\documentclass[24pt]{article}
\usepackage{amssymb}
\usepackage{amsfonts}
\usepackage{amssymb,amsmath}
\usepackage{mathrsfs}
\usepackage{latexsym}
\usepackage{amsmath}
\usepackage{latexsym}
\usepackage[cp1251]{inputenc}
\usepackage{graphicx}
\usepackage{color}

\title{Analytical representation of an iterative formula for a quantum wave impedance determination in a case of a piecewise constant potential.}
\author{O. I. Hryhorchak\\
{\small Department for Theoretical Physics, Ivan Franko National
University of Lviv,}\\
{\small 12, Drahomanov Str., Lviv, UA--79005,
Ukraine}\\
\small{\it{Orest.Hryhorchak@lnu.edu.ua}}}

\def\ch{\mathop{\rm ch}\nolimits}
\def\sh{\mathop{\rm sh}\nolimits}

\begin{document}
\renewcommand{\abstractname}{Abstract}
\maketitle

\begin{abstract}
An analytical solution for a quantum wave impedance in a case of piesewise constant potential was derived. It is in fact an analytical depiction of a well-known iterative method of a quantum wave impedance determination. The expression for a transmission probability as a function of a particle energy for an arbitrary cascad of constant potentials was obtained. The application of obtained results  was illustrated on a system of double-well/barrier structures. 
\end{abstract}

\section{Introduction}

When we talk about the approximate investigation of a quantum mechanical systems with a complicated geometry of a potential the classical approach is as follows. One depicts the real potential as a cascade of constant potentials and then applies one of the various methods for getting a solution, for example a transfer matrix formalizm \cite{Ando_Itoh:1987, Griffiths_Steinke:2001, Pereyra_Castillo:2002, SanchezSoto_atall:2012,Harwit_Harris_Kapitulnik:1986,Capasso_Mohammed_Cho:1986,Miller_etall:1985},  a finite difference method \cite{Zhou:1993, Grossmann_Roos_Stynes:2007}, a quantum wave impedance approach \cite{Kabir_Khan_Alam:1991, Hague_Khondker:1998, Nelin_Imamov:2010, Babushkin_Nelin:2011_1, Ashby:2016} and others \cite{Calecki_Palmier_Chomette:1984, Tsu_Dohler:1975, Lui_Fukuma:1986, Babushkin_Nelin:2011, Nelin_Imamov:2010}. 

Generally saying we can approximate the real potential not only by a piesewise constant potential but also by a linear, parabolic, exponential forms of a potential \cite{Arx7:2020}. It is worth to say that  these methods gives us the possibility to formalize the process of quantum systems studying. But each of them has drawbacks. For example, transfer matrix formalizm demands constructing transfer matrices for each individual area of a considered potential and then performing the operation of their multiplication. And although only two matrices for each individual area of potential are required (one describes the propagation of a wave function over a region of a constant potential and the other one relates wave functions at the interfaces of different regions of a potential) it is sometimes practically too much complicated task, especially in a case when a considered potential is approximated by a large number of intervals of constant potentials.

The other very effective method of a theoretical investigation of a quantum mechanical systems is a quantum wave impedance approach \cite{Arx1:2020, Arx2:2020, Arx3:2020, Arx4:2020, Arx5:2020}.  
In \cite{Arx3:2020} we otained the well-known iterative formula for a quantum wave impedance determination and in \cite{Arx6:2020} we discussed a numerical investigation of systems with complicated geometry of a potential using this iterative approach. 

But the iterative approach also has its drawbacks. One of them is that it does not allow an explisit analysing the dependence of a quantum wave impedance on the value of a potential energy of each interval. Thus the quaestion arises, whether it is possible to depict an iterative formula for a quantum wave impedance in  an analytical form. This problem, namely obtainingan an analytical representation of  iterative formula  is the main aim of this paper. 

\section {Analytical representation of an approximate itera\-tive calculation of a quantum wave impe\-dance}

Consecutive iterative cycles of a quantum wave impedance calculation can be depicted as the multiplication of appropriate matrices. At the end we have to multiply the result matrix on the $\begin{pmatrix} 1 \\ 1 \end{pmatrix}$ vector column. The final result for a value of a quantum wave impedance we get as a fraction of a top row on a bottom one of the result vector column. We will notify such operations with the $\rightarrow$ sign. For example the following relation
\begin{eqnarray}
Z_2=z_2\frac{z_1\ch(\gamma_2l_2)-z_2\sh(\gamma_2l_2)}
{-z_1\sh(\gamma_2l_2)+z_2\ch(\gamma_2l_2)}
\end{eqnarray}
we can depict as
\begin{eqnarray}
Z_2 \rightarrow z_2
\begin{pmatrix}
z_1\ch(\gamma_2l_2) -z_2\sh(\gamma_2l_2) 
\\
-z_1\sh(\gamma_2l_2)+ z_2\ch(\gamma_2l_2) 
\end{pmatrix}
\end{eqnarray}
or
\begin{eqnarray}
Z_2 \rightarrow z_2
\begin{pmatrix}
z_1\ch(\gamma_2l_2)& -z_2\sh(\gamma_2l_2) 
\\
-z_1\sh(\gamma_2l_2)& z_2\ch(\gamma_2l_2) 
\end{pmatrix}
\begin{pmatrix}
1 
\\
1 
\end{pmatrix},
\end{eqnarray}
and vice versa
\begin{eqnarray}
z_2
\begin{pmatrix}
z_1\ch(\gamma_2l_2)& -z_2\sh(\gamma_2l_2) 
\\
-z_1\sh(\gamma_2l_2)& z_2\ch(\gamma_2l_2) 
\end{pmatrix}
\rightarrow
z_2\frac{\!z_1\ch(\gamma_2l_2)\!-\!z_2\sh(\gamma_2l_2)}
{\!-z_1\sh(\gamma_2l_2)\!+\!z_2\ch(\gamma_2l_2)}.
\end{eqnarray}
For two consecutive regions with different characteristic impedances $z_1$, $z_2$ and different widths $l_1$, $l_2$ we have:
\begin{eqnarray}
Z_2\!\!\!\!\!\! &\rightarrow& z_2
\begin{pmatrix}
z_1\ch(\gamma_2l_2)& -z_2\sh(\gamma_2l_2) 
\\
-z_1\sh(\gamma_2l_2)& z_2\ch(\gamma_2l_2) 
\end{pmatrix}
\begin{pmatrix}
z_0\ch(\gamma_1l_1)& -z_1\sh(\gamma_1l_1) 
\\
-z_0\sh(\gamma_1l_1)& z_1\ch(\gamma_1l_1) 
\end{pmatrix}
\times
\begin{pmatrix}
1 
\\
1 
\end{pmatrix}
=\nonumber\\
\!\!\!\!\!\!\!\!&=&
\begin{pmatrix}
z_0z_1\ch(\gamma_1l_1)\ch(\gamma_2l_2)-z_1^2\sh(\gamma_1l_1)\ch(\gamma_2l_2)-\\-z_1z_2\ch(\gamma_1l_1)\sh(\gamma_2l_2)+z_0z_2\sh(\gamma_1l_1)\sh(\gamma_2l_2) 
\\
\\
z_1z_2\ch(\gamma_1l_1)\ch(\gamma_2l_2)-z_0z_2\sh(\gamma_1l_1)\ch(\gamma_2l_2)-\\
-z_0z_1\ch(\gamma_1l_1)\sh(\gamma_2l_2)+z_1^2\sh(\gamma_1l_1)\sh(\gamma_2l_2) 
\end{pmatrix}\rightarrow\nonumber\\
\!\!\!\!\!\!\!\!&\rightarrow &\!\!\!\!\!\!
z_2\frac{z_0z_1\!\ch(\gamma_1l_1)\!\ch(\gamma_2l_2)\!-\!z_1^2\!\sh(\gamma_1l_1)\!\ch(\gamma_2l_2)\!-\! z_1z_2\!\ch(\gamma_1l_1)\!\sh(\gamma_2l_2)\!+\!z_0z_2\sh(\gamma_1l_1)\!\sh(\gamma_2l_2)}
{z_1z_2\!\ch(\gamma_1l_1)\!\ch(\gamma_2l_2)\!-\!z_0z_2\sh(\gamma_1l_1)\!\ch(\gamma_2l_2)\!-\! z_0z_1\!\ch(\gamma_1l_1)\!\sh(\gamma_2l_2)\!+\!z_1^2\sh(\gamma_1l_1)\!\sh(\gamma_2l_2)}.\nonumber\\
\end{eqnarray}

For $N$ consecutive regions with characteristic impedances $z_i, i=1\ldots N$ and widthes $l_i, i=1\ldots N$ we have
\begin{eqnarray}
Z_N\rightarrow z_N\prod\limits_{i=N}^1\left\{
\begin{pmatrix}
z_{i-1}\ch(\gamma_il_i)& -z_i\sh(\gamma_il_i) 
\\
-z_{i-1}\sh(\gamma_il_i)& z_i\ch(\gamma_il_i) 
\end{pmatrix}
\right\}
\begin{pmatrix}
1 
\\
1 
\end{pmatrix}.
\end{eqnarray}
Taking into account that
\begin{eqnarray}
\ch(\gamma_il_i)=\frac{e^{\gamma_il_i}+e^{-\gamma_il_i}}{2},\quad \sh(\gamma_il_i)=\frac{e^{\gamma_il_i}-e^{-\gamma_il_i}}{2}
\end{eqnarray}
we get
\begin{eqnarray}
Z_N\rightarrow\frac{z_N}{2^N}\prod\limits_{i=N}^1\left\{
e^{\gamma_il_i}
\begin{pmatrix}
z_{i-1}& -z_i 
\\
-z_{i-1}& z_i 
\end{pmatrix}
+e^{-\gamma_il_i}
\begin{pmatrix}
z_{i-1}& z_i 
\\
z_{i-1}& z_i 
\end{pmatrix}
\right\}\!\!
\begin{pmatrix}
1 
\\
1 
\end{pmatrix}\!\!.
\end{eqnarray}
After introducing following notations 
\begin{eqnarray}
Z_j^-=\begin{pmatrix}
z_{j-1}& -z_j 
\\
-z_{j-1}& z_j 
\end{pmatrix},
\qquad
Z_j^+=\begin{pmatrix}
z_{j-1}& z_j 
\\
z_{j-1}& z_j 
\end{pmatrix}
\end{eqnarray}
we finally get the  expression for the product of $N$ matrices 
\begin{eqnarray}
\prod\limits_{j=N}^1 Z_j^{i_j}=
\begin{pmatrix}
\prod\limits_{j=1}^N (z_{j-1}+i_ji_{j-1}z_j)
\\
\prod\limits_{j=1}^N{i_N}(z_{j-1}+i_ji_{j-1}z_j)
\end{pmatrix},
\end{eqnarray}
where $i_j=\pm1$.
After simple but long transformations we get the final result
\begin{eqnarray}
Z(x_N)=Z_N=z_N\frac{\sum\limits_{\{i_j\}}K(i_j)\exp\left[-\sum\limits_{j=1}^Ni_j\gamma_jl_j\right]}
{\sum\limits_{\{i_j\}}i_NK(i_j)\exp\left[-\sum\limits_{j=1}^Ni_j\gamma_jl_j\right]},
\end{eqnarray}
where
\begin{eqnarray}
K(i_j)=\frac{1}{2^N}\prod\limits_{j=1}^N (z_{j-1}+i_ji_{j-1}z_j),
\end{eqnarray}
$z_j$ and $k_j$ are the functions of a particle energy. 
This expression can be written in the another form, namely
\begin{eqnarray}
Z(x_N)=Z_N=z_N\frac{\sum\limits_{\{i_j\}}K(i_j)\ch\left[-\sum\limits_{j=1}^Ni_j\gamma_jl_j\right]}
{\sum\limits_{\{i_j\}}i_NK(i_j)\sh\left[-\sum\limits_{j=1}^Ni_j\gamma_jl_j\right]}.
\end{eqnarray}
A condition for finding energies of bound states is as follows
\begin{eqnarray}
\frac{\sum\limits_{\{i_j\}}K(i_j)\exp\left[-\sum\limits_{j=1}^Ni_j\gamma_j\l_j\right]}
{\sum\limits_{\{i_j\}}i_NK(i_j)\exp\left[-\sum\limits_{j=1}^Ni_j\gamma_jl_j\right]}=-\frac{z_{out}}{z_N},
\end{eqnarray}
where $z_{out}$ is a characteristic impedance of the region which is on the left of a studied system.

An expression which describes the dependence of a transmission coefficient on a particle energy is as follows
\begin{eqnarray}
T(E)=1-\left|\frac{\sum\limits_{\{i_j\}}(z_N-i_Nz_0)K(i_j)\exp\left[-\sum\limits_{j=1}^Ni_j\gamma_jl_j\right]}
{\sum\limits_{\{i_j\}}(z_N+i_Nz_0)K(i_j)\exp\left[-\sum\limits_{j=1}^Ni_j\gamma_jl_j\right]}\right|^2.
\end{eqnarray}

\section{Double barrier/well systems. Analytical me\-thod}
In a paper \cite{Arx3:2020} we studied a system of two symmetric barriers/ wells using both an iterative method of a quantum wave impedance calculation and direct relations for a quantum wave impedance function. In this section we will do the same but with the help of results of the previous section.

The analytical approach which was developed in the previous section for a double barrier/well system (including nonsymmetric case) gives:
\begin{eqnarray}
Z=z_3\frac{\sum\limits_{\pm_1\pm_2\pm_3}\!\!\!\!\!
	(z_0\pm_1z_1)(z_1\pm_{1,2}z_2)(z_2\pm_{2,3}z_3)e^{\mp_1\gamma_1l_1\mp_2\gamma_2l_2\mp_3\gamma_3l_3}}
{\sum\limits_{\pm_1\pm_2\pm_3}\!\!\!\!\!\pm_3
	(z_0\pm_1z_1)(z_1\pm_{1,2}z_2)(z_2\pm_{2,3}z_3)
	e^{\mp_1\gamma_1x_1\mp_2\gamma_2x_2\mp_3\gamma_3x_3}},\nonumber\\
\end{eqnarray}
where $\pm_{1,2}=\pm_1\pm_2$, $\pm_{2,3}=\pm_2\pm_3$.
For a symmetric case we have $z_0=z_2=z$, $\gamma_0=\gamma_2=\gamma$, $z_1=z_3=\tilde{z}$, $\gamma_1=\gamma_3=\tilde{\gamma}$ and thus
\begin{eqnarray}
Z=\tilde{z}\frac{\!\!\!\!\!\sum\limits_{\pm_1\pm_2\pm_3}\!\!\!\!\!(z\pm_1\tilde{z})(\tilde{z}\pm_{1,2}z)(z\pm_{2,3}\tilde{z})e^{\mp_1\tilde{\gamma}l_1\mp_2kl_2\mp_3\tilde{\gamma}l_1}}
{\!\!\sum\limits_{\pm_1\pm_2\pm_3}\!\!\!\!\!\pm_3(z\pm_1\tilde{z})(\tilde{z}\pm_{1,2}z)(z\pm_{2,3}\tilde{z})e^{\mp_1\tilde{\gamma}l_1\mp_2kl_2\mp_3\tilde{\gamma}l_1}}\nonumber\\
\end{eqnarray}
or in the explicit form
\begin{eqnarray}
Z\!\!\!&=&\!\!\!\tilde{z}\left(\frac{}{}4\tilde{z}(\tilde{z}^2-z^2)\sh[\gamma l_2]+(\tilde{z}+z)^3e^{-\gamma l_2-2\tilde{\gamma}l_1}+(\tilde{z}-z)^3e^{-\gamma l_2+2\tilde{\gamma}l_1} -\right.\nonumber\\
&-&\left.(\tilde{z}-z)^2(\tilde{z}+z)e^{\gamma l_2-2\tilde{\gamma}l_1}-(\tilde{z}+z)^2(\tilde{z}-z)e^{\gamma l_2+2\tilde{\gamma}l_1}\right)\times\nonumber\\
&\times&\left(\frac{}{}-4z(\tilde{z}^2-z^2)\sh[\gamma l_2]+(\tilde{z}+z)^3e^{-\gamma l_2-2\tilde{\gamma}l_1}-\frac{}{}(\tilde{z}-z)^3e^{-\gamma l_2+2\tilde{\gamma}l_1}-\right.\nonumber\\
&-&(\tilde{z}-z)^2(\tilde{z}+z)e^{\gamma l_2-2\tilde{\gamma}l_1}\left.(\tilde{z}+z)^2(\tilde{z}-z)e^{\gamma l_2+2\tilde{\gamma}l_1}\right)^{-1}.
\end{eqnarray}

It is easy to transform this expression so it coincides with the expression obtained in a \cite{Arx3:2020}.

\section{Conclusions}
In this paper it was proved that the iterative process of calculation of a quantum wave impedance can be depicted analytically. So, we got the analytical solution of an equation for a quantum wave impedance with a piecewise constant potential while in  \cite{Arx3:2020} the solution was obtained in an iterative form. Analytical form of a solution is especially useful for the analysis of the relation between the geometry of a potential and an input value of a quantum wave impedance, which was demonstrated on the system of double barrier/well. The dependence of a transmission probability on a particle energy was obtained in an analytical form as well for a piesewise constant potential.

Notice, that using the analytical depiction of an iterative formula demands more computing resource for a quantum wave impedance determination that in a case of direct using iterative formula, but at the same time it gives much clear understanding of influence of each interval of a potential on the final result for a value of a quantum wave impedance. Obtained results are directly related to a design of nanodevices with the desired characteristics since they allows relating the characteristics of a device with the structure of its potential. An additional argument in favor of use a quantum wave impedance method is that it  demands ferew calculations compared to the transfer matrix approach and in many papers \cite{Kabir_Khan_Alam:1991, Hague_Khondker:1998, Vodolazka_Nelin:2013, Nelin_Liashok:2015, Nelin_Shulha_Zinher:2017, Nelin_Liashok:2016, Nelin_Shulha_Zinher:2018} it was demonstrated its efficacy for an analysis of quantum-mechanical structures with a potential which has a complicated spatial structure.

\renewcommand\baselinestretch{1.0}\selectfont


\def\name{\vspace*{-0cm}\LARGE 
	Bibliography\thispagestyle{empty}}
\addcontentsline{toc}{chapter}{Bibliography}

{\small

	\bibliographystyle{gost780u}
	\bibliography{full.bib}
	
}

\newpage

\end{document}